# Supersymmetric Solutions of D-Dimensional Dirac Equation for Woods-Saxon Potential in Minimal Length Formalism

**A Suparmi[1], J Akbar[1] and C Cari[1],**

[1]Department of Physics, Sebelas Maret University, Surakarta, Indonesia

Correspondence should be addressed to A. Suparmi; soeparmi@staff.uns.ac.id

**Abstract**. We obtain the energy eigenvalues and radial wave functions of the D-Dimensional Dirac equation in the case of spin symmetry for Woods-Saxon potential in minimal length formalism. The radial part of the D-Dimensional Dirac equation is solved by applied the supersymmetric quantum mechanics method using the Pekeris approximation to deal with the centrifugal term. The behavior of bound-state energy eigenvalues versus dimension and also quantum number is discussed for various minimal length parameters.

## 1. Introduction

The existence of minimal length (ML), when measured an object, occurs in the order of the Planck length $\ell_p = \sqrt{G\hbar/c^3} \approx 10^{-35}$ m, is predicted by the theory of quantum gravity, string theory, black-hole physics, special and general theory of relativity, etc [1,2]. In the presence of minimal length, the ordinary Heisenberg uncertainty is modified into the Generalized Uncertainty Principle (GUP). This minimal length effect makes our basic principle is changed, as a consequence, our Dirac equation must be modified too. The Dirac equation is formed by two potentials coupling which are the attractive Lorentz scalar potential $S(r)$ and the repulsive Lorentz vector potential $V(r)$. The spin symmetry condition occurs when the scalar potential was equal to vector potential $S(r) = V(r)$, while the pseudospin symmetry occurs when $S(r) = -V(r)$ [3]. With the interest in the higher dimensional field theory, the D-Dimensional Dirac equation is studied in this work in the case of spin symmetry. Some problems in higher dimensional have been solved by many authors [4–8].

In this work, the radial part of the D-Dimensional Dirac equation is solved by using the supersymmetry quantum mechanics (SUSYQM) method and Pekeris approximation for Woods-Saxon potential. The SUSYQM method used to get the energy eigenvalues and radial wave functions for $\ell = 0$ state only, therefore, we need Pekeris approximation to solve the centrifugal potential $\ell \neq 0$ states [9–11].

The standard form of Woods-Saxon potential is defined as [12]

$$V(r) = -\frac{V_0}{1 + e^{\left(\frac{r-R}{a}\right)}}, \qquad a \ll R, \qquad 0 \leq r < \infty \tag{1}$$

where $V_0$, $a$, and $R$ are interpreted as the potential depth, surface thickness, and width of potential, respectively.

## 2. Supersymmetric Quantum Mechanics (SUSYQM)

The SUSYQM model can be introduced by defining a system consisting of supersymmetric partner potential $V_+$ and $V_-$ as

$$V_\pm(x, a_j) = \phi^2(x, a_j) \pm \frac{\hbar}{\sqrt{2m}} \phi'(x, a_j) \tag{2}$$

or take another following form potential

$$V_{eff} = V_-(x, a_j) + E_0 = \phi^2(x, a_j) - \frac{\hbar}{\sqrt{2m}} \phi'(x, a_j) + E_0 \tag{3}$$

where $V_{eff}$ is called effective potential, $V_\pm$ called supersymmetric partner potential, and $\phi$ is superpotential. Then the non-relativistic energy is given as

$$E_n^{(-)} = \sum_{i=1}^{k} R(a_i) \tag{4}$$

while the remainder $R(a_i)$ can be written as

$$R(a_i) = V_+(x; a_{i-1}) - V_-(x; a_i) \tag{5}$$

we can get total non-relativistic energy in the following form

$$E_n = E_n^{(-)} + E_0 \tag{6}$$

where $E_0$ is the ground-state non-relativistic energy. To get the ground-state wave function, we can use the following form

$$\psi_0^{(-)}(x, a_0) = N \exp\left(-\frac{\sqrt{2m}}{\hbar} \int_0^x \phi(x, a_0) \, dx\right) \tag{7}$$

where $N$ is the normalization constant.

## 3. Minimal Length Formalism

The minimal length (ML) effect makes the original Heisenberg uncertainty is modified into the Generalized Uncertainty Principle (GUP) that satisfies the generalized commutation relation [13]:

$$[x_{op}, p_{op}] = i\hbar(1 + \alpha' p^2), \qquad 0 \leq \alpha' \leq 1 \tag{8}$$

where

$$\begin{aligned} x_{op} &= x \\ p_{op} &= (1 + \alpha' p^2)p \end{aligned} \tag{9}$$

where the GUP parameter $\alpha'$ is determined from a fundamental theory [1,2]. Then Equation (9) is squared on its momentum function

$$p_{op}^2 = [(1 + \alpha' p^2)p]^2 \approx (1 + 2\alpha' p^2)p^2 \tag{10}$$

The value of $(\alpha')^2$ goes to zero because the value of $\alpha'$ is very small.

## 4. Solutions of D-Dimensional Dirac Equation in ML under Woods-Saxon potential

The general form of Dirac equation for fermionic massive spin-1/2 particles moving in a repulsive Lorentz vector potential $V(r)$ and an attractive Lorentz scalar potential $S(r)$ can be written as ($c = \hbar = 1$) [14]

$$\left(\vec{\alpha} \cdot \vec{p} + \beta(M + S(\vec{r}))\right) \psi_n(\vec{r}) = (E - V(\vec{r})) \psi_n(\vec{r}) \tag{11}$$

where $E$, $M$, and $\vec{p} = -i\vec{\nabla}_D$ respectively denote the relativistic energy of the system, the mass of the fermionic particle, and D-Dimensional momentum operator, while $\vec{\alpha}$ and $\vec{\beta}$ are $4 \times 4$ Dirac matrices.

Equation (11) can be solved by the method of separation of variables, it can be solved separately in the form of radial and angle parts. Thus, the Dirac spinor can be written in the following form

$$\psi_{n\ell m}(\vec{r}) = \begin{pmatrix} \xi_{n\ell m}(\vec{r}) \\ \int_{n\ell m}(\vec{r}) \end{pmatrix} = \frac{1}{r^{\frac{D-1}{2}}} \begin{pmatrix} F_n(r) Y_m^\ell(\theta, \varphi) \\ iG_n(r) Y_m^{\bar{\ell}}(\theta, \varphi) \end{pmatrix} \tag{12}$$

where $\xi_{n\ell m}(\vec{r})$ and $\int_{n\ell m}(\vec{r})$ are upper (larger) and lower (small) component Dirac spinors, respectively. $F_n(r)$ and $G_n(r)$ are upper and lower component radial wave functions, respectively.

$Y_m^\ell(\theta,\varphi)$ and $Y_m^{\bar{\ell}}(\theta,\varphi)$ are spin and pseudospin spherical harmonics, respectively. Then Equation (12) substituted into Equation (11), gives

$$\vec{\sigma}\vec{p}\, \int_{n\ell m}(\vec{r}) = \left(-M - S(\vec{r}) + E - V(\vec{r})\right) \xi_{n\ell m}(\vec{r}) \tag{13a}$$

$$\vec{\sigma}\vec{p}\, \xi_{n\ell m}(\vec{r}) = \left(M + S(\vec{r}) + E - V(\vec{r})\right) \int_{n\ell m}(\vec{r}) \tag{13b}$$

In the case of spin symmetry $S(r) = V(r)$ and then eliminates $\int_{n\ell m}(\vec{r})$, Equation (13a) and (13b) becomes

$$\left(p^2 + \tilde{V}(\vec{r})(E + M)\right)\xi_{n\ell m}(\vec{r}) = (E^2 - M^2)\, \xi_{n\ell m}(\vec{r}) \tag{14}$$

where $V(\vec{r}) \to \dfrac{\tilde{V}(\vec{r})}{2}$.

Here, we consider the ML effect into the Dirac equation by substituted Equation (10) into Equation (14), gives

$$\left(p^2 + 2\alpha' p^4 + \tilde{V}(\vec{r})(E + M)\right)\xi_{n\ell m}(\vec{r}) = (E^2 - M^2)\, \xi_{n\ell m}(\vec{r}) \tag{15}$$

First, we consider the case corresponding to $\alpha' = 0$ [15], which gives

$$p^4 = [((E^0)^2 - M^2) - \tilde{V}(\vec{r})(E^0 + M)]^2 \tag{16}$$

where $E^0$ is the energy of minimal length. This equation then substituted into Equation (14) with substituted $\vec{p} = -i\vec{\nabla}_D$, becomes

$$(\nabla_D^2 - 2\alpha'[((E^0)^2 - M^2) - \tilde{V}(\vec{r})(E^0 + M)]^2 - \tilde{V}(\vec{r})(E + M))\xi_{n\ell m}(\vec{r}) = -(E^2 - M^2)\, \xi_{n\ell m}(\vec{r}) \tag{17}$$

Equation (17) is the Dirac equation using the minimal length formalism.

The Dirac equation can be written in any dimension in the spherical coordinate, but new coordinate need to be introduced for this D-Dimensional system, namely hyperspherical coordinates which are defined as follows [5,16]:

$$x_1 = r \cos\theta_1$$
$$x_\sigma = r \sin\theta_1 \dots \sin\theta_{\sigma-1} \cos\phi, \qquad 2 \le \sigma \le D-1 \tag{18}$$
$$x_D = r \sin\theta_1 \dots \sin\theta_{D-2} \sin\phi,$$

where $0 \le \theta_k \le \pi$, $k = 1,2,\dots D-2$, and $0 \le \phi \le 2\pi$, where $r$ is the polar variable and $\theta_1, \theta_2, \dots \theta_{D-2}, \phi$ are the hyperangles variable. The Laplacian operator in the polar coordinate $(r, \theta_1, \theta_2, \dots \theta_{D-2}, \phi)$ of $R^D$ is given as [5,16]

$$\nabla_D^2 = \frac{1}{r^{D-1}} \frac{\partial}{\partial r}\left(r^{D-1} \frac{\partial}{\partial r}\right) - \frac{L_{D-1}^2}{r^2} \tag{19}$$

where

$$L_{D-1}^2 = -\frac{1}{\sin^{D-2}\theta_{D-1}} \frac{\partial}{\partial \theta_{D-1}}\left(\sin^{D-2}\theta_{D-1} \frac{\partial}{\partial \theta_{D-1}}\right) + \frac{L_{D-2}^2}{\sin^2\theta_{D-1}} \tag{20}$$

The hyperspherical coordinates in the D-Dimensional space from Equations (19)-(20) and the wave function in Equation (12) then substituted into Equation (17), yields

$$\left(\frac{\partial^2}{\partial r^2} - \frac{[4\ell(\ell + D - 2) + (D - 1)(D - 3)]}{4r^2} - 2\alpha'[((E^0)^2 - M^2) - \tilde{V}(\vec{r})(E^0 + M)]^2 - [\tilde{V}(\vec{r})(E + M) - (E^2 - M^2)]\right) F_n(r) = 0 \tag{21}$$

where $r \in [0, \infty]$.

The Woods-Saxon Potential in Equation (1) then substituted into Equation (21), gives

$$\left(\frac{\partial^2}{\partial r^2} - O - \frac{\gamma}{r^2} - \frac{\kappa}{1 + e^{\left(\frac{r-R}{a}\right)}} - \frac{v}{\left(1 + e^{\left(\frac{r-R}{a}\right)}\right)^2}\right) F_n(r) = 0 \tag{22}$$

where

$$\gamma = \frac{[4\ell(\ell + D - 2) + (D - 1)(D - 3)]}{4}, \qquad O = 2\alpha'\left(((E^0)^2 - M^2)\right)^2 - (E^2 - M^2) \tag{23}$$

$$\kappa = V_0[4\alpha'((E^0)^2 - M^2)(E^0 + M) - (E + M)], \qquad v = 2\alpha' V_0^2 (E^0 + M)^2 \tag{24}$$

because there is a centrifugal term $\gamma/r^2$ in Equation (22), this equation cannot be solved analytically for $\ell \neq 0$ states. Therefore, we shall use the Pekeris approximation, by changing the coordinates [11]

$$x = \frac{r-R}{R}, \qquad \varpi = \frac{R}{a}, \qquad e^{\left(\frac{r-R}{a}\right)} = e^{\varpi x} \tag{25}$$

then a new form of centrifugal potential is defined as

$$V^*(x) = \frac{\gamma}{R^2}\left(C_0 + \frac{C_1}{1+e^{\varpi x}} + \frac{C_2}{(1+e^{\varpi x})^2}\right) \tag{26}$$

where $C_0$, $C_1$, and $C_2$ are new arbitrary parameters. We expand this new form of centrifugal potential $V^*(x)$ in a series around the point $x = 0$ ($r \approx R$) to get the value of $C_0$, $C_1$, and $C_2$. Then we obtain

$$C_0 = 1 - \frac{4}{\varpi} + \frac{12}{\varpi^2}, \qquad C_1 = \frac{8}{\varpi} - \frac{48}{\varpi^2}, \qquad C_2 = \frac{48}{\varpi^2} \tag{27}$$

Now, Equation (26)-(27) substituted into Equation (22), gives

$$\left(\frac{\partial^2}{\partial r^2} - O - \frac{\gamma C_0}{R^2} - \frac{\left(\kappa + \frac{\gamma C_1}{R^2}\right)}{1+e^{\left(\frac{r-R}{a}\right)}} - \frac{\left(\nu + \frac{\gamma C_2}{R^2}\right)}{\left(1+e^{\left(\frac{r-R}{a}\right)}\right)^2}\right)F_n(r) = 0 \tag{28}$$

Then Equation (28) can be solved using the SUSQM by defining the superpotential $\phi$ as follow

$$\phi(r) = -\frac{\hbar}{\sqrt{2\mu}}\left(A + \frac{B}{1+e^{\left(\frac{r-R}{a}\right)}}\right) \tag{29}$$

Substituted Equations (2) and (3) with the Equations (28) and (29), yields

$$A^2 = O + \frac{\gamma C_0}{R^2}, \quad 2AB - \frac{B}{a} = \left(\kappa + \frac{\gamma C_1}{R^2}\right), \quad B^2 + \frac{B}{a} = \left(\nu + \frac{\gamma C_2}{R^2}\right) \tag{30}$$

we solved Equation (30) by elimination and the results are

$$B = -\frac{1}{2a} + \sqrt{\nu + \frac{\gamma C_2}{R^2} + \frac{1}{4a^2}}, \qquad A = \frac{1}{-\frac{1}{a} + 2\sqrt{\nu + \frac{\gamma C_2}{R^2} + \frac{1}{4a^2}}}\left(\kappa + \frac{\gamma C_1}{R^2}\right) + \frac{1}{2a} \tag{31}$$

and

$$A = \frac{\mathcal{u}}{2B} - \frac{B}{2}, \qquad \mathcal{u} = \kappa + \nu + \frac{\gamma(C_1+C_2)}{R^2} \tag{32}$$

Next, we solved the ground-state wave function using Equation (7), and the result is

$$F_n^{(-)}(r) = N\, e^{Ar}\left(1 + e^{-\left(\frac{r-R}{a}\right)}\right)^{-aB} \tag{33}$$

with $N$ is the normalization constant. After that, we substituted Equation (29) into (2) to solve the supersymmetric partner potential, yields

$$V_\mp(r; a_0) = \frac{\hbar^2}{2\mu}\left(A^2 + \frac{B^2 \pm \frac{B}{a}}{\left(1+e^{\left(\frac{r-R}{a}\right)}\right)^2} + \frac{2AB \mp \frac{B}{a}}{1+e^{\left(\frac{r-R}{a}\right)}}\right) \tag{34}$$

If we now consider a mapping of the form

$$B \to B_1 = B_0 - \frac{1}{a}, \qquad B_n = B_0 - \frac{n}{a} \tag{35}$$

and solving the Equation (5) using Equation (32), (34), and (35) we get the following results

$$R(a_i) = -\frac{\hbar^2}{2\mu}\left[\left(\frac{\mathcal{u}}{2\left(B-\frac{i}{a}\right)} - \frac{\left(B-\frac{i}{a}\right)}{2}\right)^2 - \left(\frac{\mathcal{u}}{2\left(B-\frac{(i-1)}{a}\right)} - \frac{\left(B-\frac{(i-1)}{a}\right)}{2}\right)^2\right] \tag{36}$$

where the remainder $R(a_i)$ is independent of $r$. By solving Equation (30), we can get the ground-state relativistic energy eigenvalues equation as

$$E - M = \frac{1}{(E+M)}\left[2\alpha'(((E^0)^2 - M^2))^2 + \frac{\gamma C_0}{R^2} - A^2\right] \tag{37}$$

Then we solved Equation (6) with Equations (4), (36), and (37). The result is

$$(E^2 - M^2) = 2\alpha'\left(((E^0)^2 - M^2)\right)^2 + \frac{\gamma C_0}{R^2} - \left(\frac{u}{2\left(B - \frac{n}{a}\right)} - \frac{\left(B - \frac{n}{a}\right)}{2}\right)^2 \tag{38}$$

Equation (38) is the total relativistic energy eigenvalues equation, while the binding energy of a bound Dirac particle is defined as [17]

$$E_b = E - M \tag{39}$$

In the case of $\alpha' = 0$ (without minimal length effect), $D = 3$ and reduced to Schrödinger-like equation, then Equation (38) consistent with reference [10,18].

## 5. Results and discussion

From Equation (38) and (39) the bound-state energy eigenvalues are listed in Table 1-2 and calculated by using the following parameters: $M = 10$ fm$^{-1}$, $a = 0.5$ fm, $R = 7$ fm, $V_0 = -10$ fm$^{-1}$, $E^0 = 10$ fm$^{-1}$, and quantum number $n = 0$, for different values of minimal length parameter $\alpha'$ as a function of D-Dimensional $D$, and the azimuthal quantum number $\ell$, respectively.

The Dirac equation always produces two solutions because the energy eigenvalue $E$ has a quadratic form, $E^{upper} > E^{lower}$. In this work, we used upper eigenvalues for calculation.

**Table 1.** The bound-state energy eigenvalues for different values of minimal length parameter $\alpha'$ as a function of $D$, for $\ell = 20$

| D | $E_b$ (fm$^{-1}$) | | | |
|---|---|---|---|---|
|   | $\alpha' = 0$ | $\alpha' = 0.001$ | $\alpha' = 0.005$ | $\alpha' = 0.01$ |
| 1 | -20.15403 | -6.29935 | -1.56183 | -0.77031 |
| 2 | -20.13403 | -6.28549 | -1.54786 | -0.75579 |
| 3 | -20.11635 | -6.27129 | -1.53355 | -0.74091 |
| 4 | -20.09900 | -6.25677 | -1.51892 | -0.72570 |
| 5 | -20.08128 | -6.24192 | -1.50396 | -0.71014 |

**Table 2.** The bound-state energy eigenvalues for different values of minimal length parameter $\alpha'$ as a function of $\ell$, for $D = 3$

| $\ell$ | $E_b$ (fm$^{-1}$) | | | |
|---|---|---|---|---|
|   | $\alpha' = 0$ | $\alpha' = 0.001$ | $\alpha' = 0.005$ | $\alpha' = 0.01$ |
| 20 | -20.11635 | -6.27129 | -1.53355 | -0.74091 |
| 21 | -20.08128 | -6.24192 | -1.50396 | -0.71014 |
| 22 | -20.04353 | -6.21126 | -1.47306 | -0.67802 |
| 23 | -20.00178 | -6.17930 | -1.44088 | -0.64455 |
| 24 | -19.95546 | -6.14608 | -1.40742 | -0.60975 |

In Table 1-2, the bound-state energy eigenvalues $E_b$ have a negative value for standard Woods-Saxon potential, similar to reference [19]. From Table 1-2, the bound-state energy eigenvalues $E_b$ decrease with an increase in $D$ or $\ell$, respectively. We selected quantum number $\ell = 20$ in Table 1 and $\ell = 20$-24 in Table 2 to avoid the imaginary part that appears in bound-state energy eigenvalues results. We investigate the minimal length effect by making a variation in minimal length parameter $\alpha'$. From Table 1-2, the results show that in any arbitrary $D$ or $\ell$, the bound-state energy eigenvalues

$E_b$ decrease with an increase in minimal length parameter $\alpha'$, it means, the minimal length effect makes the bound-state energy eigenvalues decrease in any arbitrary $D$ or $\ell$. Thus, the bound-state energy eigenvalues can be decreased by an increase in D-Dimensional $D$, the azimuthal quantum number $\ell$, or the minimal length parameter $\alpha'$.

**6. Conclusion**

In this work, we have obtained the solutions of the D-Dimensional Dirac equation for the Woods-Saxon potential in minimal length formalism for the spin symmetry case. We obtain the energy eigenvalues and radial wave functions on the standard Woods-Saxon potential. Also, we have obtained the relation between the bound-state energy eigenvalues as a function of the D-Dimensional $D$, and the azimuthal quantum number $\ell$, respectively, with the variation of the minimal length parameter $\alpha'$. The results show that the bound-state energy eigenvalues can be decreased by an increase in $D$, $\ell$, or $\alpha'$. Our results in Equation (38) can be reduced to the Schrödinger-like equation and the results consistent with reference [10,18]. The results can be used to test the watched data in nuclear and high-energy physics for Woods-Saxon potential.

**Acknowledgments**

This research was partly supported by Mandatory Research Grant of Sebelas Maret University with contract number 452/UN27.21/PN/2020.

**References**
[1]  K. Konishi, G. Paffuti, and P. Provero, Phys. Lett. B **234**, 276 (1990).
[2]  J. L. Cortés and J. Gamboa, Phys. Rev. D - Part. Fields, Gravit. Cosmol. **71**, 1 (2005).
[3]  J. N. Ginocchio, Phys. Rep. **414**, 165 (2005).
[4]  D. Agboola, Pramana - J. Phys. **76**, 875 (2011).
[5]  A. Chatterjee, Phys. Rep. (1990).
[6]  S. Dong, (n.d.).
[7]  S. M. Ikhdair and R. Sever, Int. J. Mod. Phys. C **19**, 221 (2008).
[8]  S. M. Ikhdair, Chinese J. Phys. **46**, 291 (2008).
[9]  C. Berkdemir, A. Berkdemir, and R. Sever, Phys. Rev. C - Nucl. Phys. **72**, 1 (2005).
[10] V. H. Badalov, H. I. Ahmadov, and A. I. Ahmadov, Int. J. Mod. Phys. E **18**, 631 (2009).
[11] H. Rahimov, H. Nikoofard, S. Zarrinkamar, and H. Hassanabadi, Appl. Math. Comput. **219**, 4710 (2013).
[12] R. D. Woods and D. S. Saxon, Phys. Rev. **95**, 577 (1954).
[13] F. Brau, J. Phys. A. Math. Gen. **32**, 7691 (1999).
[14] G. Wei and S. Dong, Phys. Lett. A **373**, 49 (2008).
[15] M. Alimohammadi and H. Hassanabadi, Nucl. Phys. A **957**, 439 (2017).
[16] K. J. Oyewumi, Found. Phys. Lett. **18**, 75 (2005).
[17] W. Greiner, *Relativistic Quantum Mechanics* (Springer Berlin Heidelberg, 1995).
[18] C. Berkdemir, A. Berkdemir, and R. Sever, Phys. Rev. C - Nucl. Phys. **74**, 39902 (2006).
[19] V. H. Badalov, H. I. Ahmadov, and S. V. Badalov, Int. J. Mod. Phys. E **19**, 1463 (2010).